\documentclass{aastex}
\usepackage{spr-astr-addons}
\newcommand{\kms}{{\,\rm km\,s}^{-1}}

\renewcommand{\mag}{\mbox{$\;$mag}}
\makeatletter
\def\refname{References}
\renewenvironment{thebibliography}[1]{%
 \section*{\refname}%
 \thebib@list 
 \def\refpar{\relax}%
 \def\newblock{\hskip .11em\@plus.33em\@minus.07em}%
 \clubpenalty4000 
 \widowpenalty4000 
 \sfcode`\.=1000\relax 
}{%
 \endlist 
 \revtex@pageid 
}%
\makeatother
\begin{document}

\title{Allan Sandage and the Cosmic Expansion}

\epsscale{0.6}
\vspace*{-4.2cm}
\centerline{\plotone{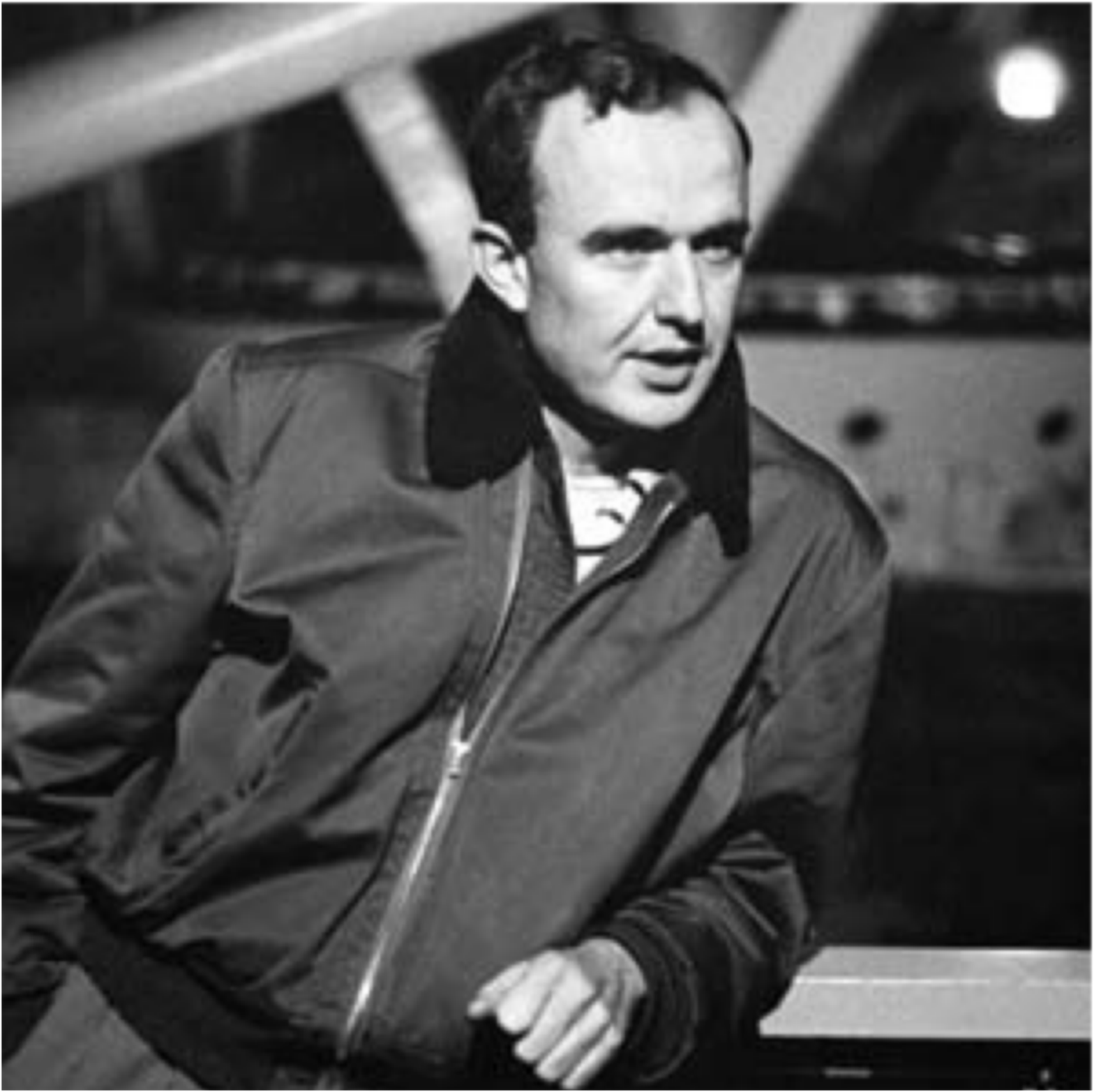}}

\author{G.~A. Tammann and B. Reindl}
\affil{%
     Department of Physics and Astronomy, University of Basel, \\
     Klingelbergstrasse 82, 4056 Basel, Switzerland}
\email{g-a.tammann@unibas.ch}

\vspace*{0.8cm}

\begin{abstract}
     This is an account of Allan Sandage's work on (1) The character
     of the expansion field. For many years he has been the strongest
     defender of an expanding Universe. He later explained the CMB
     dipole by a local velocity of  $220\pm50\kms$ toward the Virgo
     cluster and by a bulk motion of the Local supercluster (extending
     out to $\sim\!3500\kms$) of $450 - 500\kms$ toward an apex at
     $l=275$, $b=12$. Allowing for these streaming velocities he found
     linear expansion to hold down to local scales ($\sim300\kms$). 
     (2) The calibration of the Hubble constant. Probing different
     methods he finally adopted -- from Cepheid-calibrated SNe\,Ia and
     from independent RR~Lyr-calibrated TRGBs --
     $H_{0}=62.3\pm1.3\pm5.0\kms\;$Mpc$^{-1}$. 
\end{abstract}

\section{Introduction}
\label{sec:1}
\begin{figure}[t]
   \epsscale{0.96}
   \epsscale{0.98}
\plotone{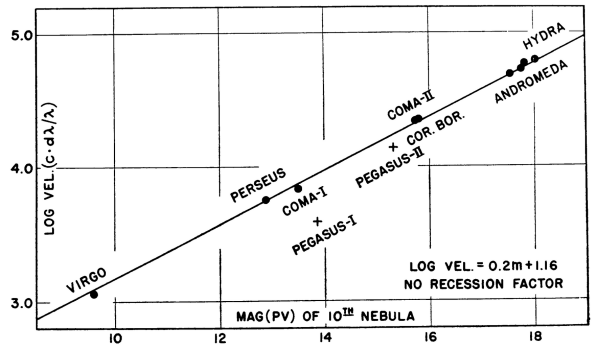}
  \caption{Hubbles last Hubble diagram with unrealistically small
    scatter and the remark no recession factor which documents his
    doubts about the expansion of the Universe.}
  \label{fig:01}
\end{figure}

Edwin \citeauthor{Hubble:29} is generally credited for the discovery
of the expansion of the Universe. 
But as so often in the case of fundamental discoveries, others had
contributed. G.~Lemaitre had published a value of the expansion rate
(Hubble constant, $H_{0}$) already in \citeyear{Lemaitre:27}, and
H.~P.~Robertson once laconically told Sandage, Hubble found the
expansion because I told him. In fact, Robertson had published his
value of $H_{0}$ already in \citeyear{Robertson:28}.
Hubbles most astounding achievement is to have convinced the World of
the expansion of the Universe with his brilliant monograph
\textit{The Realm of the Nebulae} \citeyearpar{Hubble:36a};  
he had by then much better cards than in 1929 because he had
extended with the help of Milton Humason the
$\log$ redshift-apparent magnitude diagram (Hubble diagram) to
$19,000\kms$ \citep{Hubble:Humason:34},
but his value of $H_{0}$ was still to high by a factor of roughly 8,
and correspondingly his expansion age was impossibly short -- 
a problem which he elegantly managed to bypass. 
Paradoxically \citeauthor{Hubble:36b} began to question the reality of
the expansion in the same year as his book appeared because he could
not make sense of his galaxy counts. His doubts persisted until his 
death as evidenced in his Darwin Lecture -- posthumously edited by
Sandage -- where \citeauthor{Hubble:53} showed a Hubble diagram
including \citeauthor*{Humason:51} large-redshift clusters out
$61,000\kms$ (Fig.~\ref{fig:01}) with the remark `no recession factor
(applied)', which means that he had corrected the galaxy magnitudes
for a single factor of $z$, but not for the $z^{2}$-term required in
any expanding model. 

     A definitive description of the expansion had to proceed along
two lines. 
(1) The expansion field had to be mapped in different directions and
out to truly cosmic distances -- allowing for deceleration and/or
acceleration -- to test whether the expansion is linear, which means
that it is observed as the same by any observer in the
Universe.\footnote{%
   Note: linear expansion does not require a linear Hubble line.}

(2) Only then it is meaningful to search for the cosmic value of
$H_{0}$, which in turn would provide the first cosmological test,
i.e.\ the expansion age of the Universe as compared with independent
geological and astrophysical age determinations. Sandage has
contributed to these two topics more than anybody else, although only
about one fourth of his papers are devoted to them. 
 
     The remaining possibility that redshifts are \textit{not} caused
by the cosmic expansion has been disproved later by
\citeauthor{Sandage:10} in a series of papers on the difficult Tolman
test \citep[][and references therein]{Sandage:10} which requires that
the surface brightness of a galaxy within a \textit{metric} radius
decreases with $z^{-4}$.   

\begin{figure}[t]
    \epsscale{0.80}
\plotone{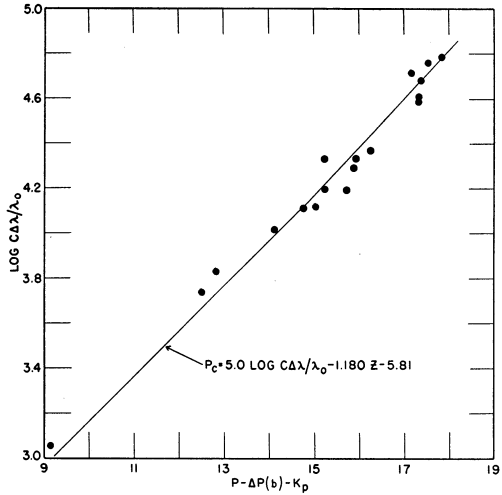}
  \caption{The Hubble diagram of first-ranked cluster galaxies from
    HMS. The curved Hubble line corresponds to a closed Universe with
    $q_{0}=2.5$.} 
  \label{fig:02}
\end{figure}

\section{The Character of the Expansion Field}
\label{sec:2}
The famous \citeauthor*{HMS:56} paper gave new support for an
isotropic, expanding Universe. M.~Humason and N.~Mayall published in
it the 630 galaxy redshifts of the combined Mount Wilson and Lick
Observatory sample. 
The task of the theoretical analysis of the data fell upon Sandage. He
homogenized the magnitudes, applied the first correct
redshift-dependent K-corrections, and he showed Hubble diagrams for
various subsamples. In particular he derived the Hubble diagram of 18 
first-ranked cluster galaxies, where he applied corrections for
luminosity evolution and the $K$-correction.
From the upwards curvature of the Hubble line he concluded that the
expansion is decelerated. In a subsequent paper \citeauthor{HS:56}
defined the deceleration parameter $q_{0}$ and derived a value of 
$q_{0} = 2.5\pm1$, i.e.\ a decelerating Universe. It is interesting
that this value flatly disagrees with $q_{0} = -1$, the value
required by Hoyle's Steady State model which he still maintained for a
long time. The \citeauthor*{HMS:56} paper was the strongest support
for an expanding Universe until 1962, when the Cosmic Microwave
Background (CMB) was detected (Fig.~\ref{fig:02}). 

     In \citeyear{Sandage:61} Sandage wrote a paper
\textit{The Ability of the 200-inch Telescope to Discriminate between
  Selected World Models}. 
It became the foundation of modern observational cosmology and made
cosmology a quantitative science. He calculated the form of the Hubble
diagram, the number of galaxies per apparent magnitude bin, and the
diameter-redshift relation for a grid of different values of $q_{0}$,
including $q_{0}=-1$.   

\begin{figure}[t]
   \epsscale{0.78}
\plotone{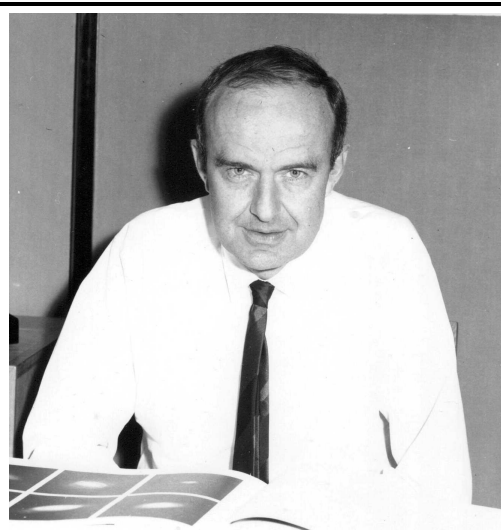}
  \caption{Allan Sandage in 1967. Under high pressure he had developed
    painful arthritis of his fingers, which is reflected in this
    picture.} 
  \label{fig:03}
\end{figure}
     The large redshifts of Quasars were discovered in 1963.
Sandage's r{\^o}le in the discovery is well \mbox{described} by
\citet{Lynden-Bell:Schweizer:11}.
Sandage also discovered the radio-quiet QSS \citep{SV:65,Sandage:65}.
The discovery prompted widespread speculations about large 
non-cosmological redshifts. Sandage was appalled. At the IAU General
Assembly in Prague in 1966 Sandage, together with Sir Martin Ryle, was
the main speaker on the nature of large redshifts. He gave a
flamboyant presentation, but some remained still unconvinced. Sandage
felt an enormous pressure for the coming years, and he developed
painful arthritis in his fingers, that later became dormant
(Fig.~\ref{fig:03}).

\subsection{The Hubble diagram of brightest cluster galaxies}
\label{sec:2:1}
\begin{figure}[t]
   \epsscale{0.82}
\plotone{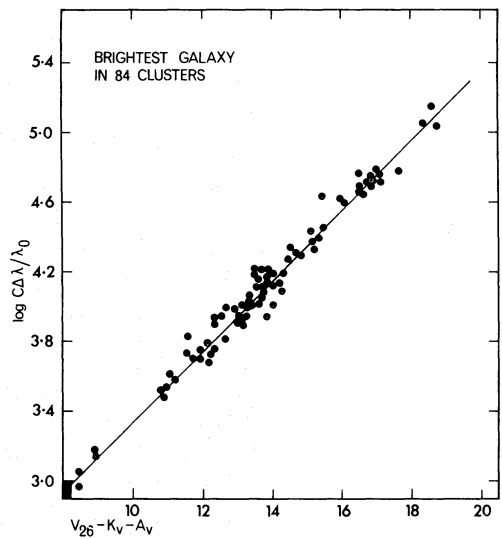}
  \caption{Sandage's Hubble diagram of first-ranked cluster galaxies
    extending to $z=0.46$. Lines for different values of $q_{0}$ are
    shown. The box in the lower left marks the interval within which
    Hubble established the redshift-distance relation in 1929.}  
  \label{fig:04}
\end{figure}

During that time Sandage decided that the Hubble diagram of brightest
cluster galaxies had to be carried to higher redshifts with the double
purpose of determining $q_{0}$ and to see how Quasars, radio galaxies,
and N and Seyfert galaxies fitted into the picture. He single-handedly
mounted a gigantic observing program for the identification, position,
apparent magnitude and redshift of distant brightest cluster galaxies
down to the limit of the 200-inch telescope. Precise positions were
needed because the fainter objects could not be seen by eye, and the
aperture photometry and photographic spectroscopy had to be done by
blind offsets. He sometimes spent 14 hours without interruption in the
narrow prime focus of the telescope, and he frequently changed,
depending on the seeing conditions, the very heavy instruments during
the night, which impaired his health. In total he invested more than
100 nights of the ``Big Eye'' on the program, that resulted in eight
papers leaving no doubt that in order to explain the scatter in the
Hubble diagram of various objects it was \textit{not} necessary to
invoke mysterious redshift, but that it was caused by the respective
luminosity functions. By \citeyear{Sandage:72} he had extended the
Hubble diagram with a dispersion of $\sim\!0.3\mag$ to $z=0.46$
as shown in Fig.~\ref{fig:04}. A formal solution for $q_{0}$ gave
$q_{0}\sim1.0\pm0.5$, yet excluding luminosity evolution.
At the arrival of CCD detectors \citeauthor{WKS:75} extended the
Hubble diagram to $z=0.75$, but without quoting a value of $q_{0}$,
because in the mean time it had become clear that the light of
E galaxies is dominated by red giants 
\citep{Baldwin:etal:73,Tinsley:73,Tammann:74} 
and that luminosity evolution has a decisive effect.

\subsection{The Hubble diagram of Supernovae of type Ia (SNe\,Ia) and
  of clusters} 
\label{sec:2:2}
Early Hubble diagrams of SNe\,I showed promise that they may become
useful as standard candles, but the dispersion was still large 
\citep{Kowal:68,Tammann:77,Tammann:78,Tammann:79,Branch:Bettis:78}
Nevertheless a scatter of less than $0.3\mag$ was suggested in a paper
by \citet{CST:85}.
The situation improved further with the spectroscopic separation of
type Ia SNe from other subtypes \citep{Branch:86}.
This led to a luminosity dispersion of $\sim\!0.25\mag$
\citep{Tammann:Leibundgut:90,Branch:Tammann:92} which made SNe\,Ia  
strong competitors to brightest cluster galaxies as standard candles,
in particular as they are presumably little affected by luminosity
evolution. Their study was followed up by many authors, too numerous
to be cited here, who increased the sample and improved the data 
\citep[e.g.][]{Hamuy:etal:96}.
The suggestion of \citet{Phillips:93} that the SN\,Ia luminosity
depended on the decline rate was initially questioned by 
\citet{TS:95}, but later fully confirmed on the basis of more distant
SNe\,Ia with reliable velocity distances in \citet{PST:00}
and \citet{RTS:05}. The dispersion of the maximum magnitude was now
reduced to $0.16\mag$, or even less for the $I$ magnitudes in
dust-free E/S0 galaxies. (For the definition of the 
corrected maximum magnitudes see \S~\ref{sec:3:3}).  
The last paper in that series \citep[][in the following SRT10]{SRT:10}
contains 246 SNe\,Ia with $v_{\rm CMB} < 30,000\kms$ (for the
corrected velocities see \S~\ref{sec:2:7}). The sample is a
compilation of $\log v$ and $m(\max)$ data from five large,
overlapping sets of SNe\,Ia (cited in \citeauthor*{SRT:10}) which were
homogenized by requiring that each set has to comply on average with
the same, arbitrarily chosen value of $H_{0}=60$. The resulting Hubble
diagram (Fig.~\ref{fig:05} below) carries Sandage's expectation of a
linearly expanding Universe down to scales of $\sim\!1200\kms$.
Others have carried the SNe\,Ia Hubble diagram to much higher
redshifts and have thereby discovered dark energy
\citep{Riess:etal:98,Perlmutter:etal:99}.  
   \epsscale{1.42}
\begin{figure*}[t]
\plotone{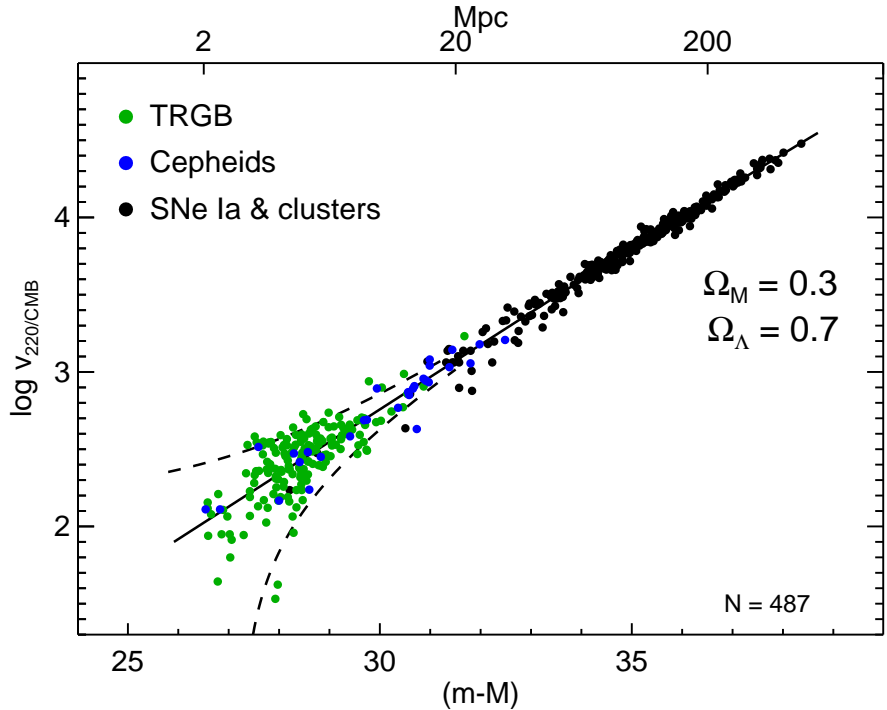}
  \caption{The composite Hubble diagram of 176 galaxies with TRGB
    distances (green) and 30 galaxies with Cepheid distances (blue);
    the 246 SNe\,Ia and 35 clusters are shown in black. The distance
    moduli $(m-M)$ are arbitrarily normalized to $H_{0}=60$. Galaxies
    with more than one distance determination are shown at the 
    mean modulus. The velocities $v_{220}$ are corrected for
    Virgocentric infall; if $v_{220}>3500\kms$ they are 
    also corrected for the motion of the Local Supercluster toward the
    corrected CMB apex (see \S~\ref{sec:2:7}). The fitted, slightly
    curved Hubble line corresponds to a $\Lambda$CDM model with
    $\Omega_{M}=0.3, \Omega_{\Lambda}=0.7$.  
    The scatter is due to distance errors \textit{and} to peculiar
    velocities. The effect of peculiar velocities of $\pm150\kms$ is
    shown by the two curved envelopes.}   
  \label{fig:05}
\end{figure*}

     In addition to the SNe\,Ia good relative distances of 28 clusters
with $3000<v< 10,000\kms$ have become available from the mean
21cm-line width distances of about 25 individual cluster members per
cluster \citep{Masters:08}. Their Hubble line has a scatter of only
$0.15\mag$ and shows no deviations from linear expansion; the line was
fitted onto the line of SNe\,Ia by a shift in apparent modulus. They
are included in Figure~\ref{fig:05}. Also fitted onto the Hubble line
are 11 clusters with good relative Fundamental-Plane (FP) distances
from \citet{Jorgensen:etal:96}.

\subsection{The Hubble diagram of Cepheids}
\label{sec:2:3}
\citet{Sandage:86} traced the Hubble diagram also to lower velocities,
using mainly Cepheid distances; the aim was to detect the perturbation
of the Local Group on the local expansion field. Later, a sample of 29
Cepheids with a minimum distance of $4.4\;$Mpc was formed from the
list of \citet{Saha:etal:06} including a few additions. This sample,
shown in Figure~\ref{fig:05} after normalization to the fiducial value
of $H_{0}=60$, defines a Hubble diagram with a dispersion of
$0.34\mag$, much of which is caused by random velocities. An
orthogonal fit to the data, assuming equal errors in magnitude and
velocity, gives a slope of $0.200\pm0.010$, i.e.\ fully consistent
with 0.2 for an isotropic Universe and $z\ll1$.

\subsection{The Hubble diagram of the tip of the red-giant branch (TRGB)}
\label{sec:2:4}
The absolute magnitude $M^{*}$ of the TRGB has emerged as a powerful
distance indicator, but, hardly reaching the Virgo cluster, its range
is still limited -- even more so than that of Cepheids. But locally
the apparent TRGB magnitudes $m^{*}$ are ideal to trace the \textit{mean}
Hubble line because their large number compensates for the large
scatter in $\log v$ caused by the random velocities of nearby field
galaxies. $m^{*}$ magnitudes of 176 galaxies have been compiled
\citep[][in the following TSR\,08a]{TSR:08a} of which the nearer ones
may be affected by the perturbation of the Local Group.
The Hubble line with only the 78 more distant ones with $m^{*}>28.2$
has a slope of $0.199\pm0.019$ in agreement with linear expansion. The
sample of 176 TRGB is shown in Figure~\ref{fig:05}, adjusted to the
fiducial value of $H_{0}=60$.  

     For the nearby Cepheid and TRGB distances it is important to note
that all distances in Figure~\ref{fig:05} refer to the barycenter of
the Local Group assumed to lie at two thirds of the way toward M\,31
(\citeauthor*{TSR:08a}).

\subsection{A composite Hubble diagram}
\label{sec:2:5}
The Hubble diagrams of SNe\,Ia (including 35 clusters), of the Cepheid
distances, and of the TRGB magnitudes have been combined in a single
diagram in Figure~\ref{fig:05} on the assumption that they comply to a
common value of $H_{0}$. The question is to what extent the assumption
is justified. 

     The intercept of the Hubble line of the SNe\,Ia has an error of
$\sigma(\log v)=0.004$. The corresponding error of the Cepheid Hubble
line is 0.012. Hence the two partially overlapping Hubble lines can be
connected within an error of $\epsilon(\log v)=0.013$ or
$\pm0.07\mag$. -- The 78 TRGB galaxies outside $4.4\;$Mpc determine
the intercept within $\sigma(\log v)=0.007$; merging them with the
SNe\,Ia and Cepheid data causes hence an additional error of
$\epsilon(\log v)=0.010$ or $\pm0.05\mag$. The combined fitting error
between the nearest and the most distant objects is therefore
$\epsilon(\log v)=0.016$ or $0.08\mag$. 
This limits the variation of $H_{0}$ with distance to about $\pm4\%$.
This value is now independent of any a priori assumption on $H_{0}$.    

     Additional evidence for the near constancy of $H_{0}$ over the
entire distance range comes from Table~\ref{tab:01} below, where the
value of $H_{0}$ of the distant SNe\,Ia is given as well as the
\textit{independent} value of the nearby TRGB distances, including
their statistical errors. From this follows a difference of $H_{0}$ of
only $1\pm4\%$.  

     The conclusion is that the cosmic value of $H_{0}$ is the same as
the mean local value at $\sim\!300\kms$ to within $\la4\%$.

\subsection{Tests of various distance indicators against linear expansion}
\label{sec:2:6}
\begin{figure}[t]
   \epsscale{0.86}
\plotone{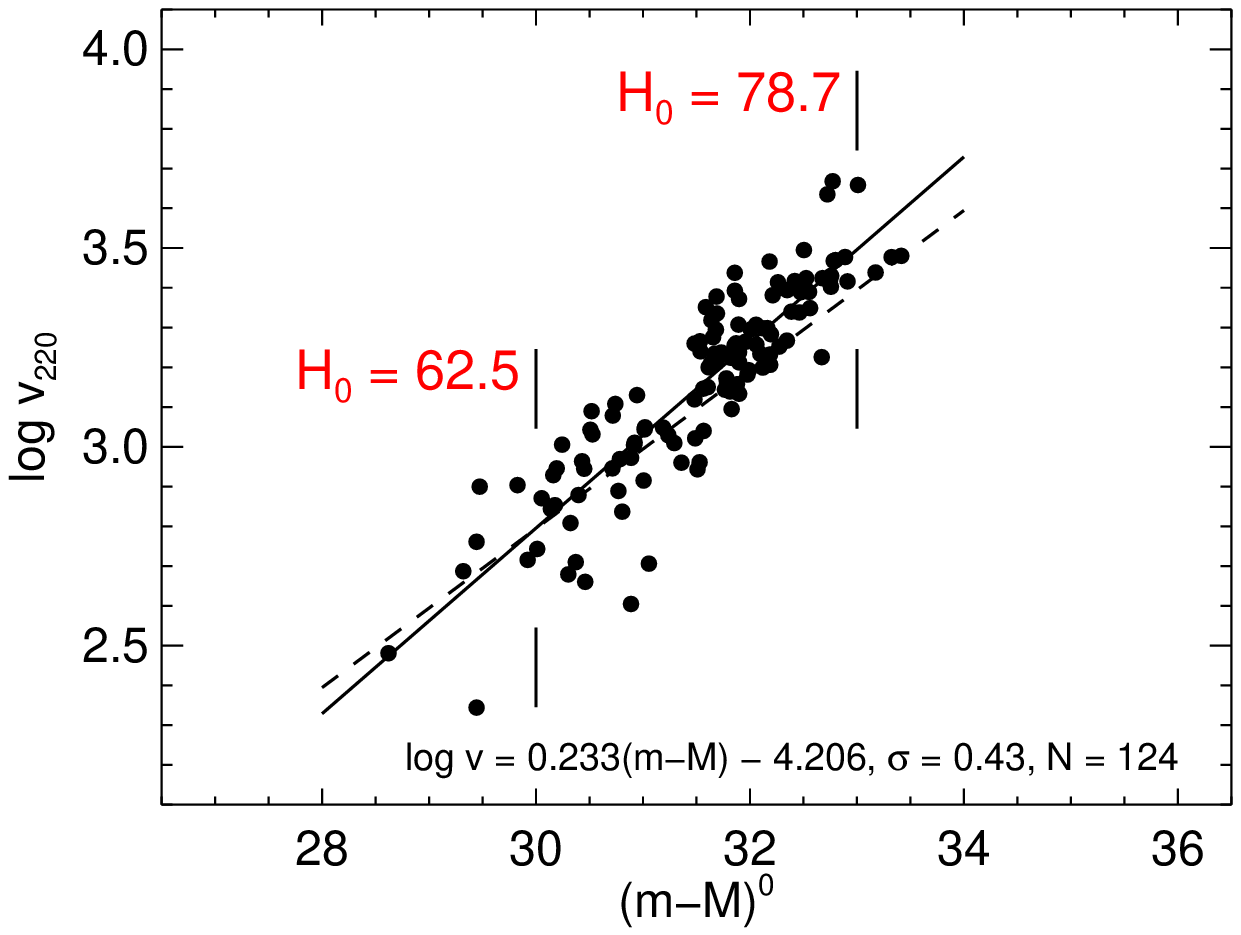}
\plotone{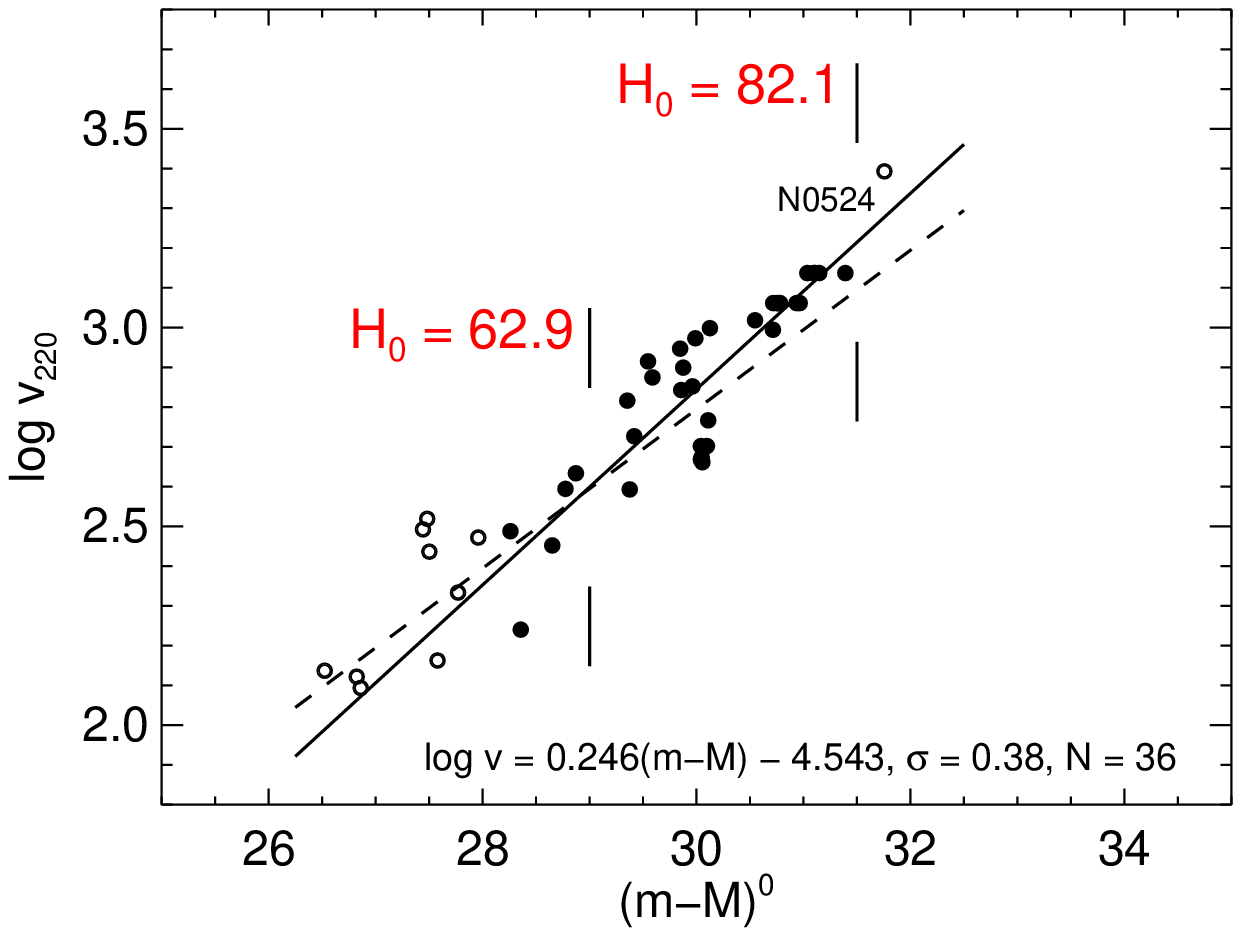}
  \caption{a) The Hubble diagram from SBF distances;
    data from \citet{Tonry:etal:01}.
    b) The Hubble diagram from PNLF distances; data compiled from
    \citet{Ciardullo:etal:02,Feldmeier:etal:07,Herrmann:etal:08}.
    The dashed line stands for linear expansion.}  
  \label{fig:06}
\end{figure}

The linearity of the expansion allows to test the results of various
distance indicators which -- beyond $300\kms$ -- must yield mean
values of $H_{0}$ that are independent of distance. Examples are: the
distances from surface brightness fluctuations (SBF) collected in
\citet{Tonry:etal:01} and the luminosity function of planetary nebulae
(PNLF) \citep[e.g.][]{Ciardullo:etal:02,Feldmeier:etal:07,Herrmann:etal:08}.
As seen in Figure~\ref{fig:06} they suggest that $H_{0}$ increases
beyond $500\kms$ by more than 25\% which is impossible in the light of 
Figure~\ref{fig:05}. However, new work on the SBF method is promising;
in any case the Fornax cluster modulus of $31.54\pm0.02$
\citep{Blakeslee:11} is in good agreement with Sandage's value of
$31.62\pm0.10$ (\citeauthor*{TSR:08a}).   

     A large sample of relative D$_{n}-\sigma$ distances of early-type
galaxies out to $10,000\kms$ has been published by \citet{Faber:etal:89}.
The sample is not complete in any sense and yields a Hubble diagram
with a scatter of $0.7\mag$. The corresponding incompleteness bias
causes a seeming, but spurious increase of $H_{0}$ with distance.
The authors have therefore applied a bias correction which causes
$H_{0}$ to \textit{decrease} by 10\% out to the catalog limit, which
suggests that the sample was somewhat overcorrected
(\citeauthor*{SRT:10}, Fig.~3).
-- A much smaller sample of related FP distances is apparently
bias-free; it has been used in \S~\ref{sec:2:2}.

     21cm line width distances (Tully-Fisher relation) of inclined
spiral galaxies have been determined by numerous authors. The crux of
the method is its large intrinsic scatter of $\sim\!0.7\mag$, that is
partially due to the difficult corrections for inclination and
internal absorption. (The apparent scatter of
\textit{magnitude}-limited samples is of course smaller.)
Distance determinations of field galaxies by some authors are
therefore affected by incompleteness bias.  

     An attempt to correct 21cm line distances for bias does not prove,
but is consistent with linear expansion \citep{FST:94}. A complete,
distance-limited and therefore bias-free sample of 104 inclined field
spirals can be defined out to only $\sim\!1000\kms$
(\citeauthor*{TSR:08a}); its large scatter does not allow to test for
linearity. Useful, however, are the nearly complete spiral samples of
the Virgo and UMa clusters (\citeauthor*{TSR:08a}). 

     The valuable cluster distances derived from many 21cm line width
data of a incomplete, but carefully bias-corrected sample of spiral
members \citep{Masters:08} are mentioned already in \S~\ref{sec:2:2}.  

\begin{figure}[t]
   \epsscale{0.99}
\plotone{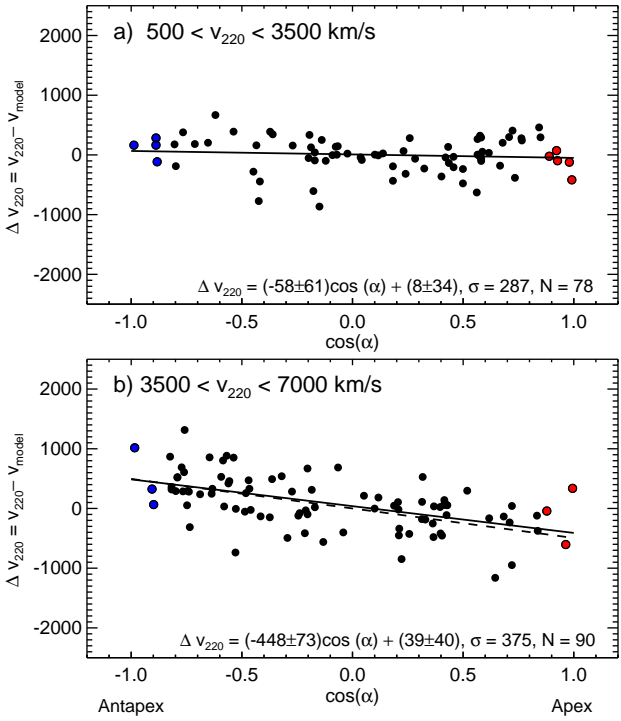}
  \caption{a) The velocity residuals $\Delta v_{220}$ versus
    $\cos\alpha$, where $\alpha$ is the angle between the object and the
    corrected CMB apex $A_{\rm corr}$. The nearly horizontal line
    indicates that nearby objects with $500<v_{220}<3500\kms$ have no
    significant systematic motion toward $A_{\rm corr}$.
    b) Same as a), but for objects with $3500<v_{220}<7000\kms$.  
    The slanted line indicates a bulk motion of the Local Supercluster
    of $448\pm73\kms$ with respect to the Machian frame.
    Red (blue) points lie within $30^{\circ}$ of the corrected apex
    (antapex).}  
  \label{fig:07}
\end{figure}

\subsection{The local dipole velocity field}
\label{sec:2:7}
%
\subsubsection{The Virgocentric infall vector of the Local Group}
\label{sec:2:7:1}
The first models of the velocity perturbations caused by the nearby
Virgo cluster are due to \citet{Silk:74} and \citet{Peebles:76}.
Sandage and some of his collab\-orators authored several papers on the
subject (e.g. \citealt{YST:80}; \citealt{ST:82a};
\citealt{Kraan-Korteweg:85}; \citealt{JT:93}). 
Their value of the Virgocentric infall vector of the Local Group of
$220\pm50\kms$ \citep{TS:85} encompasses most subsequent 
determinations. The value has been used to correct all velocities for
a self-consistent Virgocentric infall model, which assumes a Virgo
density profile of $r^{-2}$ and, correspondingly, that the infall of
individual galaxies scales with $r^{-1}$. An equation for the
corrected velocities $v_{220}$ is given in \citet{STS:06}.

\begin{figure}[t]
   \epsscale{0.75}
\plotone{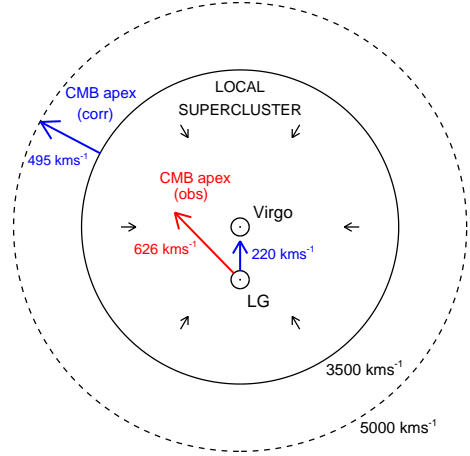}
  \caption{Schematic presentation of the local dipole velocity field.}  
  \label{fig:08}
\end{figure}

\subsubsection{The motion relative to the Cosmic Microwave Background (CMB)}
\label{sec:2:7:2}
The observed velocity of the Local Group toward the CMB apex is the
vector sum of the Virgocentric infall and of a still larger velocity
comprising a volume of unknown size. Taking the observed CMB velocity
of $626\pm30\kms$ \citep{Hinshaw:etal:07} toward an apex $A_{\rm obs}$
at $l=263.9$, $b=48.2$, reducing it to the barycenter of the Local
Group, and subtracting the Virgocentric infall one finds
$v_{\rm CMB}=495\pm25\kms$ towards the apex $A_{\rm corr}$ at $l=275\pm2$,
$b=12\pm4$. In order to determine the size of the co-moving volume a
Hubble diagram was constructed as in Fig.~\ref{fig:05}, but now using
the $v_{220}$ velocities as ordinate. The residuals $\Delta v_{220}$
from the resulting Hubble line were plotted versus $\cos\alpha$, where
$\alpha$ is the angle between the object and $A_{\rm corr}$. After
several trials the plot was divided into objects with
$v_{220}<3500\kms$ (Fig.~\ref{fig:07}a) and into objects with 
$3500 < v_{220} < 7000\kms$ (Fig.~\ref{fig:07}b). The residuals of the
nearer objects show essentially no dependence on $\cos\alpha$. This
means that they are at rest in first approximation relative to the
Virgo cluster, once the Virgocentric velocities and the regular Hubble
flow are subtracted. The inner volume emerges as the (truly
contracting) Local Supercluster (Fig.~\ref{fig:08}).
The objects with $v_{220}>3500\kms$
show a highly significant dependence on $\cos\alpha$. This reflects a
bulk motion of the Local Supercluster of $448\pm73\kms$ in good
agreement with the expected CMB value of $495\pm25\kms$. Most of the
bulk motion must therefore be caused by the gravitational force,
integrated over the whole sky, from the irregularly distributed masses
between 3500 and $<7000\kms$ (\citeauthor*{SRT:10}).

     All velocities in this paper are corrected for Virgocentric infall
and -- in case of $v_{220}>3500\kms$ -- for the adopted velocity of
$495\kms$ of the Local Supercluster toward the CMB apex $A_{\rm corr}$.

\section{The Calibration of \boldmath{$H_{0}$}}
\label{sec:3}
Hubble had based his galaxy distances on a few Cepheids, on brightest
stars, and on the mean luminosity of galaxies. His result was
$H_{0}=525$. Improvements of this value came slowly
\citep[for reviews see e.g.][]{Sandage:95,Sandage:98,Sandage:99,Tammann:06}.
In \citeyear{Baade:48} Baade defined $H_{0}$ as one of the prime
targets for the new $200''$ telescope. But his seminal distinction
between the young Population I and the old Population II
\citep{Baade:52} was still based on observations with the $100''$
telescope. The new finding, that revealed the luminosity difference
between RR~Lyr stars and Cepheids, was confirmed by Sandage's thesis
work \citeyearpar{Sandage:53} and reduced $H_{0}$ by a factor of 2.

\subsection{Sandage's work on the calibration of $H_{0}$}
\label{sec:3:1}
In \citeyear{Sandage:54} Sandage  summarized the results from the
first four years with the $200''$ telescope and concluded, mainly from
a corrected magnitude scale, that $125<H_{0}<276$ [$\kms\;$Mpc$-1$].
He also found that some of Hubble's brightest stars are actually HII
regions which are 2 magnitudes brighter; this and a new Cepheid
distance of M\,31 \citep{Baade:Swope:54} led to $H_{0}=180$
(\citeauthor*{HMS:56}). 
In \citeyear{Sandage:62} Sandage gave a review of $H_{0}$ at the
influential Santa Barbara Colloquium where he gave $H_{0}=100$ as the
mean of several authors, but his preferred value, considering also the
size of HII regions, was $H_{0}=75$.

     His well-known paper of \citeyear{Sandage:70} \textit{The search
for two numbers} ($H_{0}$ and $q_{0}$) started a new attack on $H_{0}$. 
It had begun already with the Cepheid distance of NGC\,2403
\citep{TS:68}, the first galaxy outside the Local Group, and continued
with a series of \textit{Steps toward the Hubble constant} which used
\citet{vandenBergh:60} luminosity classes of spirals in addition to
the previous distance indicators.  
The result was $H_{0}=57\pm3$ (\citealt{ST:75} and references therein). 
This prompted a 10-year controversy with G.~de~Vaucouleurs
(\citeyear{deVaucouleurs:77}, and references therein) who had
embraced a value of $H_{0}\sim100$. Subsequent papers of the series
used also 21cm line widths and the luminosity function of globular
clusters giving, if anything, somewhat lower values \citep{ST:95}.
\citet{Sandage:88} derived from the old method of the luminosity
classification of spirals a value of 42 which amused him because of
the coincidence with ``The Answer to the Ultimate Question'' in
Douglas Adams's fiction \textit{The Hitchhiker's Guide to the Galaxy}.

     After a pilot program to calibrate the luminosity of SNe\,Ia with
brightest stars \citep{ST:82b}, Sandage formed a small team to observe
with \textit{HST} the Cepheids in galaxies with known SNe\,Ia. 
Previous attempts of a SN\,Ia calibration depended mainly on an
adopted Virgo cluster distance \citep[e.g.][]{Leibundgut:Tammann:90},
which itself is controversial. The program required -- as described in
the next three Sections -- a re-evaluation of Cepheids as distance
indicators, the luminosity calibration of SNe\,Ia, and the zero-point
determination of the TRGB distances as an independent test.

\subsection{Cepheids}
\label{sec:3:2}
%
\subsubsection{The P-C and P-L relations of Cepheids}
\label{sec:3:2:1}
Sandage wrote about 50 papers on Cepheids. Already the first paper
\citeyearpar{Sandage:58} brought a new physical understanding of the
period-luminosity (P-L) relation of Cepheids which he derived from the
theory of harmonic oscillations. He showed that the P-L relation must
have intrinsic scatter, and that the relation is actually
a period-luminosity-color relation.

     A new P-L relation was constructed by superimposing the Cepheids
of several external galaxies and by setting the zero point by means of
up to 11 Cepheids that are members of Galactic clusters with known
distances \citep{ST:68,ST:69,ST:71}.

     A basic observational fact is that the colors of Cepheids depend
on metallicity. This was first set out for the Galaxy and SMC by
\citet{Gascoigne:Kron:65} and explained by \citet{Laney:Stobie:86} not
so much as a line blanketing effect, but as a temperature effect. The
metallicity effect between Galactic, LMC, and SMC Cepheids becomes
striking in their $(B\!-\!V)$ versus $(V\!-\!I)$ diagrams
\citep[][Fig.~7a\&b, in the following TSR\,03]{TSR:03}. A detailed
analysis of model atmospheres reveals that the whole instability strip
is shifted in the HR diagram by variations of the metal content
\citep{SBT:99}. If the ensuing period-color (P-C) relations are
different then the pulsation equation requires that also the P-L
relations must necessarily be metal-dependent \citep{ST:08}.
Metal-specific P-C and P-L relations are therefore needed.  

     Only for three galaxies the necessary input data, i.e.\ intrinsic
color and distance, are available:  
the Galaxy with [O/H]$_{\rm T_{e}} = 8.62$,
       LMC with [O/H]$_{\rm T_{e}} = 8.36$, and
       SMC with [O/H]$_{\rm T_{e}} = 7.98$.
The Galactic Cepheid colors are well determined
(\citealt{Fernie:etal:95}, \citeauthor*{TSR:03});
those in LMC and SMC have been derived in fields surrounding the
Cepheids and independently of the Cepheids themselves
\citep{Udalski:etal:99a,Udalski:etal:99b}.
The distances of LMC [$(m-M)=18.52$] and SMC [$(m-M)=18.93$] are known
to better than $\pm0.10\mag$ from a number of distance indicators that
are independent of any assumption on the P-L relation of Cepheids
(\citeauthor*{TSR:08b}, Table~6 \& 7). The Galactic P-L relation
relies on 33 Cepheids in Galactic clusters and associations and on 36
Cepheids with Baade-Becker-Wesselink distances; for the individual
sources see \citet{STR:04}. The two methods have been criticized by
\citet{vanLeeuwen:etal:07}, and the BBW method is blemished by the
uncertain projection factor $p$ \citep{Nardetto:11}. Yet the steep
slopes of the Galactic P-L relation from the independent cluster
Cepheids and the BBW method \citep{Fouque:etal:03} agree exceedingly
well, and the steep slope is also observed in the metal-rich galaxies
NGC\,3351 and 4321 (\citeauthor*{TSR:08b}).  

\begin{figure}[t]
   \epsscale{1.00}
\plotone{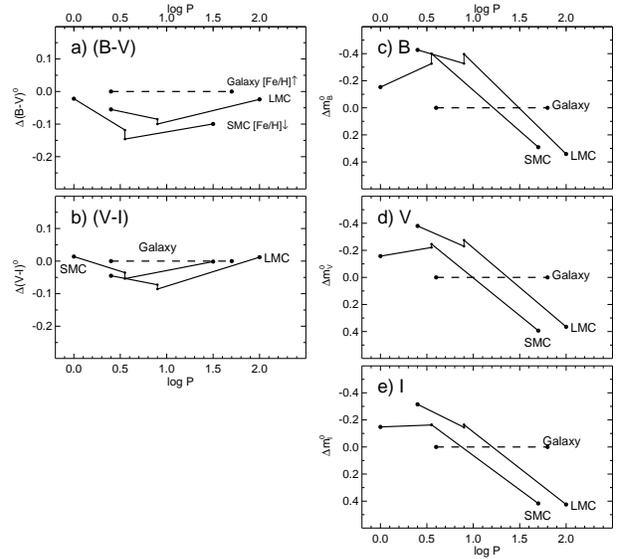}
  \caption{a) and b) The P-C relations in $(B\!-\!V)$ and $(V\!-\!I)$
    of the Cepheids in the very metal-poor SMC and in the metal-poor
    LMC \textit{relative} to the metal-rich Solar neighborhood. c)-e)
    The P-L relations in $B$, $V$, and $I$ of the Cepheids in SMC and
    LMC \textit{relative} to the Solar neighborhood.}  
  \label{fig:09}
\end{figure}

     The finally adopted, only slightly revised P-C und P-L relations
of the three calibrating galaxies are spelled out in Sandage's last
paper (\citeauthor*{TRS:11}). The relations of LMC and SMC with their
conspicuous breaks at $\log P=0.55$ and $0.9$, respectively, are
compared here with the Galactic ones in Figure~\ref{fig:09}.

     Cepheids in five galaxies of very low metallicity like SMC, or
even lower, yield particularly well to the application of the SMC P-C
and P-L relations. The resulting distances agree with RR~Lyr star and
TRGB moduli to within $\le0.05\mag$ on average. This provides an
interesting comparison of the independent distance scales of the young
Population I and old Population II.  

     It has been proposed to use so-called Wesenheit pseudo-magnitudes
$\omega$ in order to deal with the problem of internal absorption.
They are defined as 
$\omega_{V}=m_{V} - R_{V}(B\!-\!V)$ or
$\omega_{I}=m_{I} - R_{I}(V\!-\!I)$, 
where $R_{\lambda}$ is the absorption-to-reddening ratio.
\textit{Intrinsic} color differences of Cepheids with different
metallicity are treated here -- after multiplication with
$R_{\lambda}$! -- as an absorption effect. This leads of course to
systematic distance errors.

\subsubsection{Difficulties with Cepheids}
\label{sec:3:2:2}
The crux of Cepheid distances is that the internal absorption must be
known which necessitates a priori assumptions about their
(metal-dependent!) colors. Three problematic cases are mentioned in
the following.\\

\noindent
\textbf{M101.}
The 28 Cepheids in an outer metal-poor field of M\,101
\citep{Kelson:etal:96} give with the adopted P-C and P-L relations of
LMC a small internal reddening of $E(V\!-\!I)=0.03$ and
$(m-M)^{0}=29.28\pm0.05$.
The 773 Cepheids (after exclusion of overtone pulsators) in two inner,
metal-rich fields \citep{Shappee:Stanek:11} must be compared with the
metal-rich P-C relation of the Galaxy resulting in excesses
$E(V\!-\!I)$ that increase with period. The absorption-corrected P-L
relation, however, is significantly flatter than the Galactic P-L
relation, but agrees well -- in spite of higher metallicity -- with
the one of LMC. If the latter is adopted the modulus becomes
$29.14\pm0.01$.
Both of the two discrepant distance determinations are internally
consistent inasmuch as either fulfills the important test that the
\textit{individual} Cepheid distances must not depend on the period.
It seems to follow that the inner, metal-rich Cepheids are more
luminous than assumed, or that the metal-poor, outer Cepheids are
bluer and consequentially more absorbed than assumed.\\

\noindent
\textbf{NGC 4258.}
\citet{Macri:etal:06} have provided 34 Cepheids in an outer, metal-poor  
field of NGC\,4258 and 84 Cepheids in an inner, presumably metal-rich  
field. \citet{Diaz:etal:00} and \citet{Kudritzky:11}, however, have
shown that the inner field is almost as metal-poor as the outer field.
The Cepheids in both fields should therefore be reduced with the P-C
and P-L relations of LMC. One obtains then for the outer 
field $E(V\!-\!I)=0.03\pm0.03$ and $(m-M)^{0}=29.47\pm0.02$
and for the inner field $E(V\!-\!I)=0.13\pm0.05$
and $(m-M)=29.18\pm0.02$. The modulus discrepancy of $\sim\!0.3\mag$
is worrisome. The Cepheids in the two fields, although of similar
metallicity, do not seem to follow identical P-C and/or P-L
relations.

     It has been proposed to use NGC\,4258 as a cornerstone for the
distance scale because of its water maser distance of $29.29\pm0.09$
\citep{Herrnstein:etal:99} and in spite of its remaining error.
However, for other Cepheids, even if metal-poor, it is not clear
whether they should be compared with the Cepheids in the outer or
inner field.\\ 
 
\begin{figure}[t]
   \epsscale{0.88}
\plotone{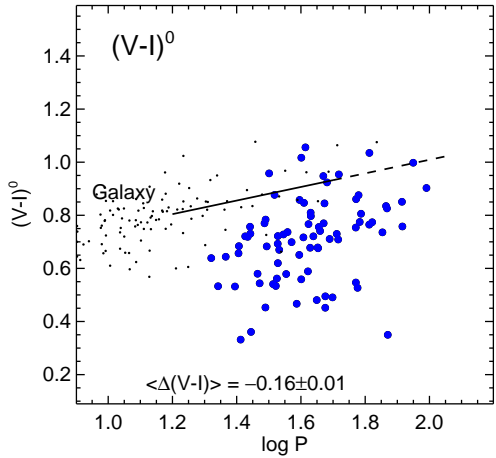}
  \caption{The $(V\!-\!I)$ colors of Cepheids in NGC\,1309.}  
  \label{fig:10}
\end{figure}

\noindent
\textbf{Blue Cepheids.} 
The metal-rich Cepheids of NGC1309 \citep{Riess:etal:09} have a P-C  
relation with unusually large scatter and are in $(V\!-\!I)$, even
without a reddening correction, $0.16\mag$ bluer on average than the
presumably equally metal-rich Galactic Cepheids (Fig.~\ref{fig:10}).
In fact they are by far the bluest long-period Cepheids known. 
The effect went unnoticed because of the use of the Wesenheit
pseudomagnitudes. The Cepheids constitute a new class. Without
knowledge of their true P-C and P-L relations it is of course not
possible to determine their distances. The case is alarming because
also the Cepheids of NGC\,3021 \citep{Riess:etal:09} appear to be too
blue by $0.07\pm0.03$, and additional intrinsically blue Cepheids may
appear red because of reddening in their parent galaxies. 

     These examples and particularly the ultra-blue Cepheids in
NGC\,1309 suggest that an additional, hidden parameter influences the
properties of Cepheids. It has been discussed in the literature
whether the Helium content could be the culprit 
\citep[e.g.][]{Marconi:etal:05,Bono:etal:08}.

     More recently infrared $H$-magnitudes of Cepheids in a few
galaxies have become available. They are less sensitive to absorption
and metal lines, but this does not prove them to be free of other
effects. Additional data are needed for an independent test.

\subsection{The luminosity calibration of SNe\,Ia}
\label{sec:3:3}
Different authors have homogenized SN\,Ia data in different ways.
The particulars of the method of Sandage's team are laid out in
\citet{RTS:05}. In short, their sample excludes known
spectroscopically peculiar SNe\,Ia. The SN colors $(B\!-\!V)$ and
$(V\!-\!I)$, corrected for Galactic reddening, are defined as the
difference of the $K$-corrected magnitudes $m_{B}^{\max}$,
$m_{V}^{\max}$, and $m_{I}^{\max}$. The intrinsic colors $(B\!-\!V)$
and $(V\!-\!I)$ as well as the color $(B\!-\!V)^{35}$, 35 days after
$B$ maximum, are determined from (dust-free) SNe\,Ia in E, S0 galaxies
and from outlying SNe\,Ia in spirals with a slight dependence on
$\Delta m_{15}$. Corresponding corrections for internal absorption are
applied throughout adopting a reddening-to-absorption ratio of
$R_{B}=3.65$ as required by the data (instead of the canonical value
of 4.1). The decline rate $\Delta m_{15}$ is defined as usual as the
brightness decline in magnitudes over the first 15 days after $B$
maximum. The corrected colors, normalized to $\Delta m_{15}=1.1$
become $(B\!-\!V)=-0.02$, $(V\!-\!I)=-0.27$, and $(B\!-\!V)^{35}=1.11$.

     Also the absolute magnitudes based on velocity distances show a
pronounced dependence on $\Delta m_{15}$. The additional dependence on
galaxian type disappears when the magnitudes are normalized to
$\Delta m_{15}=1.1$. The 62 SNe\,Ia, corrected for Galactic and
internal absorption and normalized to $\Delta m_{15}=1.1$, in the well
populated range of the Hubble diagram between 3000 and $20,000\kms$
have mean absolute magnitudes of $M_{B}=-19.57$, $M_{V}=-19.55$, and
$M_{I}=-19.28$ as judged from their velocity distances assuming
$H_{0}=60$. The statistical error of the mean absolute magnitudes is
only $0.02\mag$.  

     The \textit{HST} Supernova Project \citep{STS:06} gives for ten
SNe\,Ia with metallicity-corrected Cepheid distances weighted
luminosities of $M_{B}=-19.49\pm0.07$, $M_{V}=-19.46\pm0.07$, and
$M_{I}=-19.22\pm0.06$ in the system of \citet{RTS:05}. These values
compared with those in the previous paragraph yield a mean value of
$H_{0}=62.3\pm1.3$. The statistical error depends almost entirely on
the calibration and not on the definition of the Hubble line.
Correspondingly the systematic error of $\pm5$ (estimated) is
dominated by errors of the Cepheid distances. 

     To emphasize the difference between the SN magnitudes as defined
here and those used by other authors it is noted that, for instance, the
apparent SN magnitudes as reduced by \citet{Jha:etal:07} are
fainter by $\Delta m_{V}=0.13\mag$ on average than here. This is purely
the result of the definition of the corrected value of $m_{\max}$.

\subsection{The calibration of the tip of the red-giant branch}
\label{sec:3:4}
The fascinating property of the TRGB is that its calibration is
straightforward and that the maximum brightness of red giants is
limited by basic physics. Particularly stable is the near-infrared
maximum magnitude $I^{*}$ of red giants in old, metal-poor halo
populations \citep{DaCosta:Armandroff:90}, where also internal
absorption poses a minimum problem. The practical difficulty is the
observational determination of the upper limit $I^{*}$, which requires
a sufficiently large sample \textit{and} the separation of AGB stars.
For the history and model calculations of the TRGB see \citet{Salaris:11}.

The obvious way to calibrate the TRGB is by RR~Lyr stars. Sandage has
devoted 50 papers to these stars, exploring their classification,
metal content, evolution, age etc. His last metal-dependent,
non-linear luminosity calibration is 
$M_{V}({\rm RR}) = 1.109 + 0.600\mbox{[Fe/H]} + 0.140\mbox{[Fe/H]}^{2}$, i.e.
$M_{V}({\rm RR}) = 0.52\mag$ at [Fe/H]$=-1.5$ 
\citep{ST:06}. This calibration has been applied to 24 galaxies for
which RR~Lyr magnitudes are available in the literature as well as
TRGB magnitudes $I^{*}$ (for the many original sources see 
\citeauthor*{TSR:08b}). The combination of the RR~Lyr moduli with the
corresponding apparent $I^{*}$ magnitudes yields the absolute
magnitudes $M^{*}_{I}$. The mean magnitude of the sample -- with a
mean color of $(V\!-\!I)^{*}=1.6$ or [Fe/H]$=-1.5$ and omitting
two deviating cases -- is $M^{*}_{I} = -4.05\pm0.02$, where the
dispersion is $0.08\mag$ (\citeauthor*{TSR:08b}).
Exactly the same value has been found by \citet{Sakai:etal:04} from
globular cluster distances, and by \citet{Rizzi:etal:07} from fitting
the Horizontal Branch (HB) of five galaxies to a metal-corrected HB
with a known trigonometric parallax. Also the model luminosities of
\citet{Bergbusch:VandenBerg:01} and \citet{Salaris:11} are close to
the empirical calibration. 

     The question to what extent $M^{*}_{I}$ depends on the
metallicity has repeatedly been discussed in the literature. Most
authors agree that the luminosity does not change by more than
$\pm0.05\mag$ over the relevant range of $1.4<(V\!-\!I)^{*} < 1.8$
or $-2.0<\mbox{[Fe/H]}<-1.2$ (see Fig.~1 in \citeauthor*{TSR:08b}).

     The adopted TRGB moduli of 17 galaxies, for which also Cepheid
moduli are available (listed in \citeauthor*{TSR:08a}), reveal that
they are larger by a marginal amount of $0.05\pm0.03$, than the
Cepheid moduli. This shows that Sandage's TRGB and Cepheid distances,
although fully independent, are in satisfactory agreement. The
dispersion of the differences of $\sigma=0.13\mag$ suggests that the
individual TRGB and Cepheid distances carry random errors of less than
$\sim\!0.1\mag$. 

     The mean $M^{*}_{I}$ magnitudes of 240 galaxies
of the many values in the literature have been averaged and normalized
to the above calibration. The resulting distances are listed in
\citeauthor*{TSR:08a}. The subsample of 78 galaxies more distant than
$4.5\;$Mpc gives $H_{0}=62.9\pm1.6$.

     In the future it will be important to extend the range of TRGB
distances beyond $1000\kms$ in order to tie them even tighter to the
cosmic expansion field and/or to provide an independent luminosity 
calibration of SNe\,Ia.
First attempts have been made (\citeauthor*{TSR:08b};
\citealt{Mould:Sakai:09}).

\begin{deluxetable}{lrrcc}
\tablewidth{239pt}
\tabletypesize{\footnotesize}
\tablecaption{The final value of $H_{0}$.\label{tab:01}}
\tablehead{
 \colhead{Method}           &
 \colhead{$v_{\rm med}$}    &
 \colhead{$N$}              &
 \colhead{$H_{0}$}          &
 \colhead{Ref.}      
}
\startdata
TRGB                             &  350 &  78 & $62.9\pm1.6$ & 1 \\
21cm line width                  &  750 & 104 & $59.0\pm1.9$ & 2 \\
Cepheids                         &  900 &  29 & $63.4\pm1.8$ & 1 \\
SNe\,Ia ($v_{220}\!<\!2000$)     & 1350 &  20 & $60.2\pm2.7$ & 2 \\
SNe\,Ia ($v_{\rm CMB}\!>\!3000$) & 7700 &  62 & $62.3\pm1.3$ & 3 \\
\noalign{\smallskip}
\tableline
\noalign{\smallskip}
adopted           & & & \multicolumn{2}{l}{$62.3\pm1.3(\pm5.0)$} \\
\enddata
\tablenotetext{\,}{References --
   (1)\,\citeauthor*{TSR:08a};
   (2)\,\citeauthor*{TSR:08b};
   (3)\,\citealt{STS:06}. 
}
\end{deluxetable}

\subsection{Sandage's last value of the Hubble constant}
\label{sec:3:5}
Sandage has persued the calibration of $H_{0}$ for almost 60 years. It
was his aim from the beginning to base his distance scale on two
independent pillars, i.e.\ on Population~I and Population~II objects,
and he spent about equal efforts on either route. The distance scale
of the former depends heavily on Cepheids, whereas that of the
Population~II relies mainly on RR~Lyr stars. The determination of
Cepheid distances has become more involved because of the metal
dependence of the P-C and P-L relations, accentuated by the
corresponding problems of internal absorption and other unexplained
effects -- in particular of the more metal-rich Cepheids (see
\S~\ref{sec:3:2:2}). Hence the need for a second pillar has become even
more urgent. The direct comparison of Cepheids and RR~Lyr stars is
unprofitable because of the paucity of galaxies with reliable data on
both distance indicators. But here the RR Lyr-calibrated TRGB
distances jump in, which offer ample comparison with Cepheid distances
(\citeauthor*{TSR:08a}, Tab.~9). 
More important yet was for him that the -- admittedly still local --
value of $H_{0}$ from the TRGB is the same within the statistical
errors as that from Cepheids and Cepheid-calibrated 21cm distances and
SNe\,Ia as summarized in Table~\ref{tab:01}.   

     Some months before Sandage's death \citet{Reid:etal:10} published
a paper combining the catalog of luminous red galaxies (LRG) from the
Sloan Digital Sky Survey DR7 with the 5-year WMAP data and the Hubble
diagram of the SNe\,Ia Union Sample to find a value of
$H_{0}=65.6\pm2.5$ on the assumption of a $\Lambda$CDM model.  

\begin{figure}[h!]
   \epsscale{0.76}
 \plotone{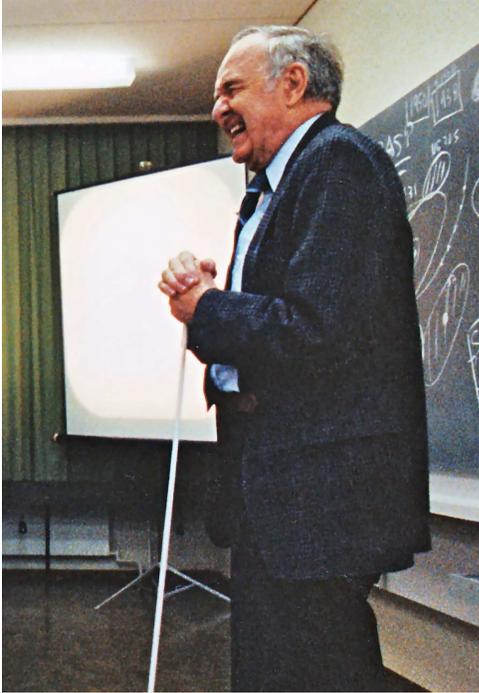}
  \caption{Allan Sandage during a lecture in Basel in 1994.}  
  \label{fig:11}
\end{figure}

\begin{acknowledgements}
A personal note of GAT: I think of Allan Sandage in deep thankfulness.
He was my inspiring mentor, he shared his love for astronomy with me,
he impressed me with his broad culture, and beyond this he was a
friend of profound honesty.  
\end{acknowledgements}




\begin{thebibliography}{}
%
\bibitem[Baade(1948)]{Baade:48}
   Baade, W.: 1948,
   \pasp\ \textbf{60}, 230 (1948)
%
\bibitem[Baade(1952)]{Baade:52}
   Baade, W.:
   Trans. IAU \textbf{8}, 397 (1952)
%
\bibitem[Baade \& Swope(1954)]{Baade:Swope:54}
   Baade, W., Swope, H.H.:
   Yearbook Carnegie Inst. \textbf{53}, 20 (1954)
%
\bibitem[Baldwin et~al.(1973)]{Baldwin:etal:73}
   Baldwin, J.R., Danziger, I.J., Frogel, J.A, Persson, S.E.:
   Astrophys. Let. \textbf{14}, 1 (1973)
%
\bibitem[Bergbusch \& VandenBerg(2001)]{Bergbusch:VandenBerg:01}
   Bergbusch, P.A., VandenBerg, D.A.:
   \apj\ \textbf{556}, 322 (2001)
%
\bibitem[Blakeslee(2011)]{Blakeslee:11}
   Blakeslee, J.P.: this Conference (2011)
%
\bibitem[Bono et~al.(2008)]{Bono:etal:08}
   Bono, G., Caputo, F., Fiorentino, G., Marconi, M., Musella, I.:
   \apj\ \textbf{684}, 102 (2008)
%
\bibitem[Branch(1986)]{Branch:86}
   Branch, D.:
   \apj\ \textbf{300}, 51 (1986)
%
\bibitem[Branch \& Bettis(1978)]{Branch:Bettis:78}
   Branch, D., Bettis, C.:
   \aj\ \textbf{83}, 224 (1978)
%
\bibitem[Branch \& Tammann(1992)]{Branch:Tammann:92}
   Branch, D., Tammann, G.A.:
   \araa\ \textbf{30}, 359 (1992)
%
\bibitem[Cadonau, Sandage, \& Tammann(1985)Cadonau et~al.]{CST:85} 
   Cadonau, R., Sandage, A., Tammann, G.A.:
   In: Bartel, N. (ed.)
   Lecture Notes in Physics \textbf{224}:
   Supernovae as Distance Indicators,
   Berlin: Springer,
   p.~151 (1985)
%
\bibitem[Ciardullo et~al.(2002)]{Ciardullo:etal:02}
   Ciardullo, R., Feldmeier, J.J., Jacoby, G.H., Kuzio de Naray, R., 
   Laychak, M.B., Durrell, P.R.:
   \apj\ \textbf{577}, 31 (2002)
%
\bibitem[Da~Costa \& Armandroff(1990)]{DaCosta:Armandroff:90}
   Da~Costa, G.S., Armandroff, T.E.:
   \aj\ \textbf{100}, 162 (1990)
%
\bibitem[de Vaucouleurs(1977)]{deVaucouleurs:77}
   de Vaucouleurs, G.:
   In: Balkowski, C., Westerlund, B.E. (eds.)
   IAU Coll. \textbf{37}:
   D{\'e}calage vers le rouge et expansion de l'univers,    
   Paris: Ed. CNRS,
   p.~301 (1977)
%
\bibitem[D{\'i}az et~al.(2000)]{Diaz:etal:00}
   D{\'i}az, A.I., Castellanos, M., Terlevich, E., Luisa Garc{\'i}a-Vargas, M.:
   \mnras\ \textbf{318}, 462 (2000)
%
\bibitem[Faber et~al.(1989)]{Faber:etal:89}
   Faber, S.M., Wegner, G., Burstein, D., et~al.:
   \apjs\ \textbf{69}, 763 (1989)
%
\bibitem[Federspiel et~al.(1994)]{FST:94}
   Federspiel, M., Sandage, A., Tammann, G.A.:
   \apj\ \textbf{430}, 29 (1994)
%
\bibitem[Feldmeier et~al.(2007)]{Feldmeier:etal:07}
   Feldmeier, J.J., Jacoby, G.H., Philips, M.M.:
   \apj\ \textbf{657}, 76 (2007)
%
\bibitem[Fernie et~al.(1995)]{Fernie:etal:95}
   Fernie, J.D., Beattie, B., Evans, N.R., Seager, S.:
   IBVS \textbf{4148} (1995)
%
\bibitem[Fouqu{\'e} et~al.(2003)]{Fouque:etal:03}
   Fouqu{\'e}, P., Storm, J., Gieren, W.: 
   Lect. Notes Phys. \textbf{635}, 21 (2003)
%
\bibitem[Gascoigne \& Kron(1965)]{Gascoigne:Kron:65}
   Gascoigne, S.C.B., Kron, G.E.:
   \mnras\ \textbf{130}, 333 (1965)
%
\bibitem[Hamuy et~al.(1996)]{Hamuy:etal:96}
   Hamuy, M., Phillips, M.M., Suntzeff, N.B., et~al.:
   \aj\ \textbf{112}, 2408 (1996)
%
\bibitem[Herrmann et~al.(2008)]{Herrmann:etal:08}
   Herrmann, K.A., Ciardullo, R., Feldmeier, J.J., Vinciguerra, M.:
   \apj\ \textbf{683}, 630 (2008)
%
\bibitem[Herrnstein et~al.(1999)]{Herrnstein:etal:99}
   Herrnstein, J.R., Moran, J.M., Greenhill, L.J.,et~al.: 
   \nat\ \textbf{400}, 539 (1999)
%
\bibitem[Hinshaw et~al.(2007)]{Hinshaw:etal:07}
   Hinshaw, G., Nolta, M.R., Bennett, C.L., et~al.: 
   \apjs\ \textbf{170}, 288 (2007)
%
\bibitem[Hoyle and Sandage(1956)]{HS:56}
   Hoyle, F., Sandage, A.:
   \pasp\ \textbf{68}, 301 (1956)
%
\bibitem[Hubble(1929)]{Hubble:29}
   Hubble, E.: 
   Proc. Nat. Acad. Sci. \textbf{15}, 168 (1929)
%
\bibitem[Hubble(1936)]{Hubble:36a}
   Hubble, E.: 
   The Realm of the Nebulae.
   New Haven: Yale Univ. Press (1936a)
%
\bibitem[Hubble(1936)]{Hubble:36b}
   Hubble, E.:
   \apj\ \textbf{84}, 517 (1936b)
%
\bibitem[Hubble(1953)]{Hubble:53}
   Hubble, E.: 
   \mnras \textbf{113}, 658 (1953)
%
\bibitem[Hubble \& Humason(1934)Humason]{Hubble:Humason:34}
   Hubble, E., Humason, M.L.:
   Proc. Nat. Acad. Sci. \textbf{20}, 264 (1934)
%
\bibitem[Humason(1951)Humason's]{Humason:51}
   Humason, M.L.:
   \pasp\ \textbf{63}, 232 (1951)
%
\bibitem[Humason et~al.(1956)HMS]{HMS:56}
   Humason, M.L., Mayall, N.U., Sandage, A.R.:
   \aj\ \textbf{61}, 97 (1956) (HMS)
%
\bibitem[Jerjen \& Tammann(1993)]{JT:93}
   Jerjen, H., Tammann, G.A.:
   \aap\ \textbf{276}, 1 (1993)
%
\bibitem[Jha et~al.(2007)]{Jha:etal:07}
   Jha, S., Riess, A.G., Kirshner, R.P.:
   \apj\ \textbf{659}, 122 (2007)
%
\bibitem[J{\o}rgensen et~al.(1996)]{Jorgensen:etal:96}
   J{\o}rgensen, I., Franx, M., Kjaergaard, P.:
   \mnras\ \textbf{280}, 167 (1996)
%
\bibitem[Kelson et~al.(1996)]{Kelson:etal:96}
   Kelson, D.D., Illingworth, G.D., Freedman, W.F., et~al.: 
   \apj\ \textbf{463}, 26 (1996)
%
\bibitem[Kowal(1968)]{Kowal:68}
   Kowal, C.T.:
   \aj\ \textbf{73}, 1021 (1968)
%
\bibitem[Kraan-Korteweg(1985)]{Kraan-Korteweg:85}
   Kraan-Korteweg, R.C.:
   In: Richter, O.G., Binggeli, B. (eds.) 
   ESO Workshop on the Virgo Cluster, 
   Garching: ESO,
   p.~397 (1985)
%
\bibitem[Kudritzky(2011)]{Kudritzky:11}
   Kudritzky, R.: this Conference (2011)
%
\bibitem[Laney \& Stobie(1986)]{Laney:Stobie:86}
   Laney, C.D., Stobie, R.S.: 
   \mnras\ \textbf{222}, 449 (1986) 
%
\bibitem[Leibundgut \& Tammann(1990)]{Leibundgut:Tammann:90}
   Leibundgut, B., Tammann, G.A.: 
   \aap\ \textbf{230}, 81 (1990)
%
\bibitem[Lema{\^i}tre(1927)]{Lemaitre:27}
   Lema{\^i}tre, G.:
   Ann. Soc. Sci. Bruxelles A. \textbf{47}, 49 (1927)
%
\bibitem[Lynden-Bell and Schweizer(2011)]{Lynden-Bell:Schweizer:11}
   Lynden-Bell, D., Schweizer, F.:
   %
   Biographical Memoirs of the Royal Society, in press (2011)
%
\bibitem[Macri et~al.(2006)]{Macri:etal:06}
   Macri, L.M., Stanek, K.Z., Bersier, D., Greenhill, L.J., Reid, M.J.: 
   \apj\ \textbf{652}, 1133 (2006) 
%
\bibitem[Marconi et~al.(2005)]{Marconi:etal:05}
   Marconi, M., Musella, I., Fiorentino, G.:
   \apj\ \textbf{632}, 590 (2005)
%
\bibitem[Masters(2008)]{Masters:08}
   Masters, K.L.:
   In: Bridle, A.H., Condon, J.J., Hunt, G.C. (eds.)
   ASP Conf. Ser. \textbf{395}:
   Frontiers of Astrophysics,
   San Francisco: ASP,
   p.~137 (2008)
%
\bibitem[Mould \& Sakai(2009)]{Mould:Sakai:09}
   Mould, J., Sakai, S.:
   \apj\ \textbf{697}, 996 (2009)
%
\bibitem[Nardetto(2011)]{Nardetto:11}
   Nardetto, N.: this Conference (2011)
%
\bibitem[Parodi et~al.(2000)]{PST:00}
   Parodi, B.R., Saha, A., Sandage, A., Tammann, G.A.:
   \apj\ \textbf{540}, 634 (2000)
%
\bibitem[Peebles(1976)]{Peebles:76}
   Peebles, P.J.E.:
   \apj\ \textbf{205}, 318 (1976)
%
\bibitem[Perlmutter et al.(1999)]{Perlmutter:etal:99}
   Perlmutter, S., Aldering, G., Goldhaber, G., et al.: 
   \apj\ \textbf{517}, 565 (1999)
%
\bibitem[Phillips(1993)]{Phillips:93}
   Phillips, M.M.:
   \apj\ \textbf{413}, L105 (1993)
%
\bibitem[Reid et~al.(2010)]{Reid:etal:10}
   Reid, B.A., Percival, W.J., Eisenstein, D.J., et~al.:
   \mnras\ \textbf{404}, 60 (2010)
%
\bibitem[Reindl et~al.(2005)]{RTS:05}
   Reindl, B., Tammann, G.A., Sandage, A., Saha, A.:
   \apj\ \textbf{624}, 532 (2005)
%
\bibitem[Riess et~al.(1998)]{Riess:etal:98} 
   Riess, A.G., Filippenko, A.V., Challis, P., et~al.: 
   \aj\ \textbf{116}, 1009 (1998)
%
\bibitem[Riess et~al.(2009)]{Riess:etal:09}
   Riess, A.G., Macri, L., Li, W., et~al.: 
   \apjs\ \textbf{183}, 109 (2009)
%
\bibitem[Rizzi et~al.(2007)]{Rizzi:etal:07}
   Rizzi, L., Tully, R.B., Makarov, D., et~al.: 
   \apj\ \textbf{661}, 815 (2007) 
%
\bibitem[Robertson(1928)]{Robertson:28}
   Robertson, H.P.: 
   Phil. Mag. \textbf{5}, 835 (1928)
%
\bibitem[Saha et~al.(2006)]{Saha:etal:06}
   Saha, A., Thim, F., Tammann, G.A., Reindl, B., Sandage, A.:
   \apjs\ \textbf{165}, 108 (2006)
%
\bibitem[Sakai et~al.(2004)]{Sakai:etal:04}
   Sakai, S., Ferrarese, L., Kennicutt, R.C., Saha, A.: 
   \apj\ \textbf{608}, 42 (2004)
%
\bibitem[Salaris(2011)]{Salaris:11}
   Salaris, M.: this Conference (2011)
%
\bibitem[Sandage(1953)]{Sandage:53}
   Sandage, A.:
   \aj\ \textbf{58}, 61 (1953)
%
\bibitem[Sandage(1954)]{Sandage:54}
   Sandage, A.:
   \aj\ \textbf{59}, 180 (1954)
%
\bibitem[Sandage(1958)]{Sandage:58}
   Sandage, A.:
   \apj\ \textbf{127}, 513 (1958)
%
\bibitem[Sandage(1961)]{Sandage:61}
   Sandage, A.: 
   \apj\ \textbf{133}, 355 (1961)
%
\bibitem[Sandage(1962)]{Sandage:62}
   Sandage, A.:
   In: McVittie, G.C. (ed.)
   IAU Symp. \textbf{15}:
   Problems of Extra-Galactic Research, 
   p.~359 (1962) 
%
\bibitem[Sandage(1965)]{Sandage:65}
   Sandage, A.: 
   \apj\ \textbf{141}, 1560 (1965)
%
\bibitem[Sandage(1970)]{Sandage:70}
   Sandage, A.:
   Phys. Today \textbf{23}, 34 (1970)
%
\bibitem[Sandage(1972)]{Sandage:72}
   Sandage, A.:
   \apj\ \textbf{178}, 1 (1972)
%
\bibitem[Sandage(1986)]{Sandage:86}
   Sandage, A.:
   \apj\ \textbf{307}, 1 (1986)
%
\bibitem[Sandage(1988)]{Sandage:88}
   Sandage, A.:
   \apj\ \textbf{331}, 583 (1988)
%
\bibitem[Sandage(1995)]{Sandage:95}
   Sandage, A.:
   In: Binggeli, B., Buser, R. (eds.)
   Saas-Fee Adv. Course \textbf{23}:
   The Deep Universe, 
   p.~1 (1995) 
%
\bibitem[Sandage(1998)]{Sandage:98}
   Sandage, A.:
   In: Livio, M., Fall, S.M., Madau, P. (eds.) 
   STScI Symp. \textbf{11}:
   The Hubble Deep Field, 
   p.~1 (1998)
%
\bibitem[Sandage(1999)]{Sandage:99}
   Sandage, A.:
   \araa\ \textbf{37}, 445 (1999)
%
\bibitem[Sandage(2010)]{Sandage:10}
   Sandage, A.:
   \aj\ \textbf{139}, 728 (2010)
%
\bibitem[Sandage et~al.(1999)SBT\,99]{SBT:99}
   Sandage, A., Bell, R.A., Tripicco, M.J.:
   \apj\ \textbf{522}, 250 (1999)
%
\bibitem[Sandage et~al.(2010)SRT10]{SRT:10}
   Sandage, A., Reindl, B., Tammann, G.A.:
   \apj\ \textbf{714}, 1441 (2010) (SRT\,10)
%
\bibitem[Sandage \& Tammann(1968)ST\,68]{ST:68}
   Sandage, A., Tammann, G.A.: 
   \apj\ \textbf{151}, 531 (1968)
%
\bibitem[Sandage \& Tammann(1969)ST\,69]{ST:69}
   Sandage, A., Tammann, G.A.: 
   \apj\ \textbf{157}, 683 (1969)
%
\bibitem[Sandage \& Tammann(1971)ST\,71]{ST:71}
   Sandage, A., Tammann, G.A.: 
   \apj\ \textbf{167}, 293 (1971)
%
\bibitem[Sandage \& Tammann(1975)ST\,75]{ST:75}
   Sandage, A., Tammann, G.A.:
   \apj\ \textbf{197}, 265 (1975)
%
\bibitem[Sandage \& Tammann(1982)ST\,82a]{ST:82a}
   Sandage, A., Tammann, G.A.:
   In: Br{\"u}ck, H., Coyne, G., Longair, M. (eds.)
   Astrophysical Cosmology: 
   Study Week on Cosmology and Fundamental Physics, 
   p.~23 (1982a)
%
\bibitem[Sandage \& Tammann(1982b)ST\,82b]{ST:82b}
   Sandage, A., Tammann, G.A.: 
   \apj\ \textbf{256}, 339 (1982b)
%
\bibitem[Sandage \& Tammann(1995)ST\,95]{ST:95}
   Sandage, A., Tammann, G.A.: 
   \apj\ \textbf{446}, 1 (1995)
%
\bibitem[Sandage \& Tammann(2006)ST\,06]{ST:06}
   Sandage, A., Tammann, G.A.: 
   \araa\ \textbf{44}, 93 (2006)
%
\bibitem[Sandage \& Tammann(2008)ST\,08]{ST:08}
   Sandage, A., Tammann, G.A.: 
   \apj\ \textbf{686}, 779 (2008)
%
\bibitem[Sandage et~al.(2004)STR\,04]{STR:04}
   Sandage, A., Tammann, G.A., Reindl, B.:
   \aap\ \textbf{424}, 43 (2004)
%
\bibitem[Sandage et~al.(2006)]{STS:06}
   Sandage, A., Tammann, G.A., Saha, A., Reindl, B., Macchetto,
   F.D., Panagia, N.:
   \apj\ \textbf{653}, 843 (2006)
%
\bibitem[Sandage \& V{\'e}ron(1965)]{SV:65}
   Sandage, A., V{\'e}ron, P.:
   \apj\ \textbf{142}, 412 (1965)
%
\bibitem[Shappee \& Stanek(2011)]{Shappee:Stanek:11}
   Shappee, B.J., Stanek, K.Z.:
   \apj\ \textbf{733}, 124 (2011)
%
\bibitem[Silk(1974)]{Silk:74}
   Silk, J.:
   \apj\ \textbf{193}, 525 (1974)
%
\bibitem[Tammann(1974)]{Tammann:74}
   Tammann, G.A.:
   In: Longair, M.S. (ed.)
   IAU Symp. \textbf{63}:
   Confrontation of cosmological theories with observational data,
   Dordrecht: Reidel,
   p.~47 (1974)
%
\bibitem[Tammann(1977)]{Tammann:77}
   Tammann, G.A.:
   In: Papagiannis, M.D. (ed.)
   Ann. NY Acad. Sci. \textbf{302}:
   Eight Texas Symposium on Relativistic Astrophysics,
   p.~61 (1977)
%
\bibitem[Tammann(1978)]{Tammann:78}
   Tammann, G.A.:
   Mem. Soc. Astron. Astron. Italiana \textbf{49}, 315 (1978)
%
\bibitem[Tammann(1979)]{Tammann:79}
   Tammann, G.A.:
   In: Longair, M.S., Warner, J.W. (eds.)
   IAU Coll. \textbf{54}:
   Scientific Research with the Space Telescope,
   Washington: NASA,
   p.~263 (1979)
%
\bibitem[Tammann(2006)]{Tammann:06}
   Tammann, G.A.:
   In: R{\"o}ser, S. (ed.)
   Rev. Mod. Astron. \textbf{19}:
   The Many Facets of the Universe -- Revelations by New Instruments,
   Weinheim: Viley-VCH,
   p.~1 (2006)
%
\bibitem[Tammann \& Leibundgut(1990)]{Tammann:Leibundgut:90}
   Tammann, G.A., Leibundgut, B.: 
   \aap\ \textbf{236}, 9 (1990)
%
\bibitem[Tammann et~al.(2011)TRS\,11]{TRS:11}
   Tammann, G.A., Reindl, B., Sandage, A.: 
   \aap\ \textbf{531}, 134 (2011) (TRS\,11)
%
\bibitem[Tammann \& Sandage(1968)TS\,68]{TS:68}
   Tammann, G.A., Sandage, A.:
   \apj\ \textbf{151}, 825 (1968)
%
\bibitem[Tammann \& Sandage(1985)]{TS:85}
   Tammann, G.A., Sandage, A.:
   \apj\ \textbf{294}, 81 (1985)
%
\bibitem[Tammann \& Sandage(1995)]{TS:95}
   Tammann, G.A., Sandage, A.: 
   \apj\ \textbf{452}, 16 (1995)
%
\bibitem[Tammann, Sandage, \& Reindl(2003)TSR\,03]{TSR:03}
   Tammann, G.A., Sandage, A., Reindl, B.: 
   \aap\ \textbf{404}, 423 (2003) (TSR\,03)
%
\bibitem[Tammann et~al.(2008a)TSR\,08a]{TSR:08a}
   Tammann, G.A., Sandage, A., Reindl, B.:
   \aapr\ \textbf{15}, 289 (2008a) (TSR\,08a)
%
\bibitem[Tammann et~al.(2008b)TSR\,08b]{TSR:08b}
   Tammann, G.A., Sandage, A., Reindl, B.: 
   \apj\ \textbf{679}, 52 (2008b) (TSR\,08b)
%
\bibitem[Tinsley(1973)]{Tinsley:73}
   Tinsley, B.M.:
   \apj\ \textbf{184}, L41 (1973)
%
\bibitem[Tonry et~al.(2001)]{Tonry:etal:01}
   Tonry, J.L., Dressler, A., Blakeslee, J.P., et~al.:
   \apj\ \textbf{546}, 681 (2001)
%
\bibitem[Udalski et~al.(1999a)]{Udalski:etal:99a}
   Udalski, A., Soszynski, I., Szymanski, M., et~al.: 
   \actaa\ \textbf{49}, 223 (1999a)
%
\bibitem[Udalski et~al.(1999b)]{Udalski:etal:99b}
   Udalski, A., Soszynski, I., Szymanski, M., et~al.:
   \actaa\ \textbf{49}, 437 (1999b)
%
\bibitem[van den Bergh's(1960)]{vandenBergh:60}
   van den Bergh, S.: 
   \apj\ \textbf{131}, 215 (1960)
%
\bibitem[van Leeuwen et~al.(2007)]{vanLeeuwen:etal:07}
   van Leeuwen, F., Feast, M.W., Whitelock, P.A., Laney, C.D.:
   \mnras\ \textbf{379}, 723 (2007)  
%
\bibitem[Sandage with J.~Westphal and J.~Kristian(1975)Westphal et~al.]{WKS:75}
   Westphal, J.A., Kristian, J., Sandage, A.:
   \apj\ \textbf{197}, L95 (1975)
%
\bibitem[Yahil et~al.(1980)]{YST:80}
   Yahil, A., Sandage, A., Tammann, G.A.:
   Physica Scripta \textbf{21}, 635 (1980)

\end{thebibliography}
\end{document}